\documentstyle[prd,aps,preprint,epsfig]{revtex}

\tighten

\begin{document}
\draft
 
\pagestyle{empty}

\preprint{
\noindent
\hfill
\begin{minipage}[t]{3in}
\begin{flushright}
UH--511--1043--04\\
LBL--54494\\
February 2004
\end{flushright}
\end{minipage}
}

\title{Broken SU(3) antidecuplet for $\Theta^+$ and $\Xi_{\frac{3}{2}}$}

\author{Sandip Pakvasa$^a$ and Mahiko Suzuki$^b$}
\address{$^a$ Department of Physics and Astronomy, University of Hawaii\\
        Honolulu, Hawaii 96822 \\
$^b$ Department of Physics and Lawrence Berkeley National Laboratory\\
University of California, Berkeley, California 94720
}


\maketitle

\begin{abstract}
   If the narrow exotic baryon resonances
$\Theta^+(1540)$ and $\Xi_{\frac{3}{2}}(1862)$ are members of the 
$J^P=\frac{1}{2}^+$ antidecuplet with $N^*(1710)$, the 
octet-antidecuplet mixing is required not only by the mass 
spectrum but also by the decay pattern of $N^*(1710)$. This casts
doubt on validity of the $\Theta^+$ mass prediction by the
chiral soliton model. While all pieces of the existing experimental 
information point to a small octet-decuplet mixing, the magnitude 
of mixing required by the mass spectrum is not consistent with 
the value needed to account for the hadronic decay rates. The discrepancy 
is not resolved even after the large experimental uncertainty is 
taken into consideration. We fail to find an alternative $SU(3)$ 
assignment even with different spin-parity assignment.  When we extend 
the analysis to mixing with a higher SU(3) multiplet, we find 
one experimentally testable scenario in the case of mixing with 
a 27-plet.
\end{abstract}
\pacs{PACS number(s):11.30.Hv,13.30.Eg,12.39.Dc,12.39.Jh}
\pagestyle{plain}
\narrowtext

\setcounter{footnote}{0}
\section{Introduction}

  Two baryon states with exotic quantum numbers were recently 
reported by experimenters\cite{Nakano,CERN}. Existence of the 
isosinglet resonance $\Theta^+(1540)$ of $Y=+1$ was corroborated
by several subsequent experiment\cite{Theta}. The discovery of 
$\Theta^+$ led to search of the exotic $\Xi_{\frac{3}{2}}$ 
resonance. The NA49 Collaboration at CERN\cite{CERN} surveyed 
over a wide mass range of $\Xi\pi$ and reported a narrow resonance, 
$\Xi_{\frac{3}{2}}$ at mass 1862 MeV.  The H1 Collaboration\cite{H1}
reported discovery of an exotic charmed baryon. Despite numerous 
positive reports, it is fair to say that some theorists and 
experimentalists are still skeptical about even their existence. 
There are hints of inconsistency among the reported values of 
the mass and the width. A very recent high-resolution mass plot 
by the HERA-B 
Collaboration\cite{HERAB} showed no evidence for $\Theta^+$ nor 
$\Xi_{\frac{3}{2}}$. From the theoretical standpoint these exotic 
resonances do not contradict the basic principles of quantum 
chromodynamics. Nonetheless they do challenge our long-standing 
beliefs and practice in hadron spectroscopy.

  The search of $\Theta^+$ was motivated by the flavor $SU(3)$
extension of the Skyrme model or the chiral soliton 
model\cite{Skyrm,Witten}. In this model the baryon number is
identified with the topological quantum number associated the
hedgehog configuration interlocking spin and flavors. After 
quantum chromodynamics was introduced, the large $N_c$ picture 
was used to argue for the classical soliton
solution\cite{Witten}. The original Skyrme model was then
extended to three flavors by embedding the Skyrme solution
into the SU(2) subgroups\cite{SkyrmeSU3}. The exotic baryons 
$\Theta^+$ and $\Xi_{\frac{3}{2}}$ fit in the antidecuplet 
${\bf \overline{10}}$ of $SU(3)$\cite{SkyrmeSU3,P,Diakonov}. 
Their spin-parity is $\frac{1}{2}^+$ since 
$\overline{{\bf 10}}$ appears as the rotationally excited levels
of the $\frac{1}{2}^+$ and $\frac{3}{2}^+$ baryons. The soliton 
model describes the baryons as made of an infinite number of 
quarks in the large $N_c$ limit. Consequently the infinite
Skyrme tower emerges for the baryons. 

After the announcement of the discovery of $\Theta^+$, numerous 
studies have been made in the constituent quark
picture\cite{JW,Lipkin,Stancu,cohen}. In the constituent quark 
model the antidecuplet baryons can be realized when five quarks 
$qqqq\overline{q}$ or more form baryons.  When the five quarks 
are bound all in relative $s$-wave, the resulting baryons would 
be in the $J^P=\frac{1}{2}^-$ states. This is a general feature 
and persists in some attempts\cite{L} with the lattice 
calculation. A strong spin-orbit force must play 
a role for the lowest pentaquark states to have 
positive parity in the constituent model. Dynamical arguments 
were put forth to lower the $J^P=\frac{1}{2}^+$ 
pentaquarks\cite{JW,Lipkin}.
While both the constituent pentaquark model and the chiral 
soliton model can accommodate the ${\bf\overline{10}}$ 
representation of $SU(3)$, they are very different 
models in nature.  Experimentally, it is possible
in principle to determine the parity of $\Theta^+$ through the
angular distribution in the production process\cite{Zao}.

The $SU(3)$ extension of the chiral soliton model predicts the 
${\bf\overline{10}}$ baryons below the second excited baryon
multiplet ${\bf 35}$ of the Skyrme tower\cite{SkyrmeSU3}. The 
exotic quantum numbers of $\Theta^+$ are a good signature of the 
${\bf\overline{10}}$ multiplet. 
The prediction of the $\Theta^+$ mass in the chiral solition model 
is based on the assumption that the $N^*(1710)$ resonance is the
nonstrange member of ${\bf \overline{10}}$\cite{P,Diakonov}.   
  While both $\Theta^+$ and $\Xi_{\frac{3}{2}}$ fit nicely to
${\bf \overline{10}}$, the nearby $\frac{1}{2}^+$ octet baryon 
resonances inevitably mix with the antidecuplet through the 
symmetry breaking.  The existing data already require
mixing and contradict the assumption that $N^*(1710)$ is 
the pure ${\bf \overline{10}}$ partner of $\Theta^+$ and 
$\Xi_{\frac{3}{2}}$. 

  In this paper we make a flavor $SU(3)$ analysis of the baryon 
masses and decay branchings without using the dynamical details 
of the constituent pentaquark model nor of the chiral soliton 
model. Our results are therefore model independent and based 
mostly on the group theory. While the group-theory analysis has 
been presented for masses in very recent papers\cite{mass}, we
focus on the consistency of broken SU(3) symmetry not only among 
the masses but also between the masses and the decay rates.

  Though purely group-theoretical analyses are less model
dependent, they have an obvious disadvantage to the dynamical 
analyses\cite{Arndt} that 
incorporate the inputs of the chiral soliton model. Only when 
a sufficient number of experimental inputs exist, can we draw
definite conclusions or make interesting predictions from 
a group-theory analysis. In carrying out our analysis, we must
control the number of independent free parameters of group 
theory. As for the higher-order symmetry breaking in the mass 
spectrum, we incorporate the second-order effects only where
they are enhanced by a small mass difference between nearby 
states through representation mixing. In the decay rates, the 
dominant symmetry breaking is by far in the phase space 
corrections. Following conventional wisdom, we do not 
include symmetry breaking in the coupling constants.

 Going through this standard group-theory analysis, 
we find difficulty in accommodating $\Theta^+$ and 
$\Xi_{\frac{3}{2}}$ in the decuplet even after the 
${\bf 8}$-${\bf\overline{10}}$ mixing is taken into account. 
In our analysis we encounter a large uncertainty in the 
masses and widths of the $SU(3)$ partners of $\Theta^+$ and 
$\Xi_{\frac{3}{2}}$. The source of the uncertainty is mainly in 
the phase-shift analysis that exposed the higher baryon 
resonances mostly in the 1960's to 70's. However, even this 
large uncertainty in the masses and widths of N* resonances
fails to resolve the discrepancy we report.

  Although there is no experimental evidence for the 
${\bf 27}$-plet baryon rsonances at present, we look for a 
possible resolution with the ${\bf 27}$-${\bf\overline{10}}$ 
mixing too. We assign the exisiting resonances, normally assigned 
to ${\bf 8}$, to the nonexotic members of ${\bf 27}$ and see if 
the difficulty in the ${\bf 8}$-${\bf\overline{10}}$ mixing  
can be resolved or not and what predictions can be made 
for future experiment. This attempt turns up with one
possible resolution. 

\section{Mass spectrum}

\subsection{The octet-antidecuplet mixing}

The smallest multiplet that can accommodate both $\Theta^+(1540)$
and $\Xi_{\frac{3}{2}}$ is the antidecuplet ${\bf \overline{10}}$.
In the simple-minded constituent pentaquark model, the states of
$J^P=\frac{1}{2}^-$, namely all in $s$-wave, are more likely
lower in mass. However, it is the chiral soliton model that
triggered our interest in these exotic resonances. Furthermore,
the original idea of Skyrme incorporated in the chiral soliton
model, if it should apply to the real world, is revolutionary in
understanding of the baryon number.  Therefore we follow the
prediction of the chiral soliton model for the quantum numbers
of $\Theta^+$ in this paper and consider
the case of spin-parity equal to $\frac{1}{2}^+$ for $\Theta^+$ 
and $\Xi_{\frac{3}{2}}$.  The representation mixing is a
second-order symmetry breaking effect. Only when two mases are close,
can it be enhanced by a small mass difference to compete with or
even dominate over the first-order breaking. In the chiral 
soliton model the mixing of $\overline{10}$ with the ground-state 
octet was studied because of the mass spectrum of the 
model\cite{Arndt}. The mixing effect with the ground-state octet
is not large enough to affect the antidecuplet masses substantially. 
Among the established $\frac{1}{2}^+$
resonances, there is the socalled Roper resonance at 1440 MeV,
which is potentially more important because of its proximity 
to $N^*(1710)$ in mass. We focus here on the Roper resonance 
$N^*(1440)$ as the mixing partner of $N^*(1710)$. Then the 
excited $\Sigma$ resonances at 1660 MeV and at 1880 MeV are the 
candidates of the $Y=0$ partners. Reviews of Particle Physics 
(RPP)\cite{PDG} lists another $N^*$ at 2100 MeV and $\Sigma^*$ 
at 1770 MeV, but both are rated with ``one star'' (evidence weak). 
Furthermore $N^*(2100)$ is too far off from the mass region of 
our interest. Table I lists the $\frac{1}{2}^+$ resonances 
relevant to our analysis. These baryon resonances were discovered 
many years ago through the partial-wave analysis of meson-baryon 
scattering.  Unfortunately, the values of mass and width 
spread substantially from one experiment to another because they 
are sensitive to details of the angular distributions and the 
methods of partial-wave analysis.  Some recent analyses\cite{PSA} 
in the baryon channels after the discovery of $\Theta^+$ have  
revealed substantial uncertainties.  The RPP makes its own 
estimate of the masses and widths from the reported results 
for each resonance and lists their best fits or likely ranges 
of values as ``our estimates''.  Without a better alternative, 
we shall adopt the values of ``our estimate'' as the experimental 
values keeping in mind that uncertainties are fairly large 
in many cases.

  The assignment of $\Theta^+$ and $\Xi_{\frac{3}{2}}$ to 
${\bf\overline{10}}$ raises an immediate problem. 
The Gell-Mann-Okubo (GMO) mass formula,
\begin{equation} 
    M({\bf n},I,Y) = M_n +m_nY + m'_n[I(I+1) - \frac{1}{4}Y^2]  
                                            \label{GMO}
\end{equation}
requires the equal spacing rule between the adjacent $Y$ members 
of ${\bf \overline{10}}$. Therefore, the mass of the pure 
antidecuplet $N^*({\bf\overline{10}})$ should be at
\begin{eqnarray}
 M_{N^*(\bf\overline{10})}&=&\frac{1}{3}(M_{\Xi_{\frac{3}{2}}} 
                  + 2M_{\Theta^+}),   \nonumber \\
                       &=&  1647\; {\rm MeV}.    \label{Ntenbar}
\end{eqnarray}
The GMO formula holds very well for the ${\bf 8}$ and ${\bf 10}$ 
baryons almost up to the intra-isospin splittings ({\em i.e.,} up 
to the electromagnetic and the $u$-$d$ mass splitting). Therefore 
the prediction of Eq. (\ref{Ntenbar}) should hold with the 
accuracy of a few MeV. Since the masses of $\Theta^+$ and 
$\Xi_{\frac{2}{3}}$ are sharply determined in experiment and their 
widths are within the experimental resolution, the number on the 
right-hand side of Eq. (\ref{Ntenbar}) should have little  
uncertainty. The $N^*(1710)$ resonance therefore does not fit 
well to the partner of $\Theta^+$ and $\Xi_{\frac{3}{2}}$. 
Putting it in another way, if $N^*(1710)$ were a pure 
${\bf\overline{10}}$, $\Xi_{\frac{3}{2}}(1862)$ would be
at 2080 MeV. A simple and natural way to reconcile with 
this difficulty is to postulate that $N^*(1710)$ is not a pure 
${\bf \overline{10}}$, but contains an octet component. 
 
  If the amount of ${\bf 8}$-${\bf\overline{10}}$ mixing  
present in $N^*(1710)$ is too large, the chiral soliton model prediction 
on the $\Theta^+$ mass made by Diakonov {\em et al.} 
\cite{Diakonov} and by their predecessors\cite{P} would lose 
its basis since it assumes that $N^*(1710)$ is the pure 
${\bf \overline{10}}$ and does not include the $O(m_s^2)$
mass correction. On the other hand, one may take an optimistic 
viewpoint that the mass 1647 MeV is not so far off the lower 
end of the likely range (1680$\sim$1740 MeV) suggested for 
$N^*(1710)$ by RPP. It may be premature to jump onto the 
negative conclusion in view of the large experimental 
uncertainty. With this optimism or skepticism, let us proceed 
to make a more quantitative analysis of the mass spectrum 
incorporating the ${\bf 8}$-${\bf\overline{10}}$ mixing. 

In the presence of the ${\bf 8}$-${\bf\overline{10}}$ mixing, 
the mass eigenstates of $N^*$ and $\Sigma^*$ are defined by
\begin{eqnarray}
  |N^*(1440)\rangle &=&|{\bf 8}\rangle\cos\theta_N -
           |{\bf\overline{10}}\rangle\sin\theta_N,\nonumber\\
  |N^*(1710)\rangle &=&|{\bf\overline{10}}\rangle\cos\theta_N+
                       |{\bf 8}\rangle\sin\theta_N, \label{N} 
\end{eqnarray} 
and by the corresponding relations
\begin{eqnarray}
 |\Sigma^*(1660)\rangle &=& |{\bf 8}\rangle\cos\theta_{\Sigma}-
    |{\bf\overline{10}}\rangle\sin\theta_{\Sigma},\nonumber\\
 |\Sigma^*(1880)\rangle &=&
                |{\bf\overline{10}}\rangle\cos\theta_{\Sigma}+
                         |{\bf 8}\rangle\sin\theta_{\Sigma}.
                     \label{Sigma}
\end{eqnarray}
Let us denote the coefficients of the GMO formula 
for ${\bf 8}$ and ${\bf\overline{10}}$ as
\begin{eqnarray}
 M({\bf \overline{10}};Y)&=&M_{\overline{10}}-aY,\nonumber \\
     M({\bf 8};I,Y) &=& M_8 - bY +c[I(I+1)-\frac{1}{4}Y^2]
\end{eqnarray}
and introduce the ${\bf 8}$-${\bf\overline{\bf 10}}$ mixing 
parameter by
\begin{equation}
  \delta\equiv \langle N^*({\bf 8})|N^*({\bf\overline{10}})\rangle=  
   \langle\Sigma^*({\bf 8})|\Sigma^*({\bf\overline{10}})\rangle.
\end{equation}
Then the mixing angles $\theta_N$ and $\theta_{\Sigma}$ are related 
to these parameters by
\begin{eqnarray}
        \tan 2\theta_N &=& \frac{2\delta}{M_{\overline{10}}
        - M_8 -a+b-\frac{1}{2}c} \nonumber \\
        \tan 2\theta_{\Sigma} &=& \frac{2\delta}{M_{\overline{10}}
        - M_8-2c}, \label{mixing}
\end{eqnarray} 
and the baryon masses are expressed as  
\begin{equation}
     \left. \begin{array}{l}
   \Xi_{\frac{3}{2}}(1862) 
         = M_{\overline{10}} + a,  \\
   \Xi^*_{\frac{1}{2}} = M_8 + b + \frac{1}{2}c, \\
   \Sigma^*(1880) + \Sigma^*(1660)
    = M_{\overline{10}} +M_8 +2c, \\
   \bigl(\Sigma^*(1880) -\Sigma^*(1660)\bigr)\cos 2\theta_{\Sigma}
    = M_{\overline{10}}- M_8 -2c,  \\
   \Lambda^*(1600) = M_8,  \\
   N^*(1710)+N^*(1440) = M_{\overline{10}}+M_8-a-b+\frac{1}{2}c,\\
   \bigl((N^*(1710) - N^*(1440)\bigr)\cos 2\theta_N =
    M_{\overline{10}}-M_8-a+b-\frac{1}{2}c,  \\
   \Theta^+(1540) = M_{\overline{10}}-2a,
           \end{array} \right.
\end{equation}
where the baryons denote their masses.
   The masses of $\Theta^+$ and $\Xi_{\frac{3}{2}}$ immediately
determine $M_{\overline{10}}$ and $a$ accurately. 
$M_8$ is fixed by the mass of $\Lambda^*(1600)$ alone, albeit
with a fairly large uncertainty ($1600_{-50}^{+100}$ MeV). The 
remaining three parameters, $b$, $c$ and $\delta$, are 
determined by other four masses. There is one redundancy.
When we fit $b$ and $c$ to $N^*(1710)$ and $N^*(1440)$,
the central values of our fit are as follows:
\begin{equation}
           \left. \begin{array}{ll}
   M_{\overline{10}} \simeq 1755\; {\rm MeV},& 
     M_8 \simeq 1600\; {\rm MeV} \\
   a \simeq 107\; {\rm MeV},\;\; b \simeq 144\; {\rm MeV}, & 
   c \simeq 93\; {\rm MeV},\\
   \delta \simeq\pm 123\;{\rm MeV}.
          \end{array} \right.  \label{fit} 
\end{equation}
Alternatively, we can determine $\delta$ using $\Sigma^*(1660)$
and $\Sigma^*(1880)$. The value thus determined ($\delta\simeq
\pm 109$ MeV) is consistent with ${\pm}$123 MeV in Eq. (\ref{fit}) after 
the large uncertainty of the masses are taken into account. 
The value of $\delta\simeq \pm 123$ MeV leads to
\begin{equation}
    \tan\theta_N \simeq \pm 0.59 \;\;\;
    \tan\theta_{\Sigma}\simeq \pm 1.12.  \label{angle}
\end{equation}
We have chosen $|\tan\theta_N|<1$ and $|\tan\theta_{\Sigma}|>1$ 
since $N^*(1710)$ is dominantly ${\bf\overline{10}}$ and 
$\Sigma({\bf 8})$ (1785 MeV) is heavier than 
$\Sigma({\bf\overline{10}})$ (1755 MeV) (cf. Eq. (\ref{Sigma})).
Within the uncertainties due to the spread of the mass values 
reported by different phase-shift analyses\cite{PDG}, the fit 
shows not only that  $\tan\theta_N$ is nonzero, but also 
that the ${\bf 8}$-${\bf\overline{10}}$ mixing is substantial. 
We quote from Eq. (\ref{angle})
\begin{equation}
      \tan^2\theta_N^{(m)} \simeq 0.35, \label{m}
\end{equation} 
where the superscript of $\theta_N^{(m)}$ indicates the value 
determined by the mass spectrum. Let us compare this mixing with 
that of the ``correlated-diquark'' model by Jaffe and 
Wilczek\cite{JW}. Their $N^*(1440)$ is a linear combination of
\begin{equation}
  |N^*(1440)\rangle = \sqrt{\frac{2}{3}}|{\bf 8}\rangle +
                \sqrt{\frac{1}{3}}|{\bf \overline{10}}\rangle,
\end{equation}
namely $\tan^2\theta_N = 0.5$. This mixing is considerably 
larger than the value obtained from our analysis of the mass 
spectrum.

   The set of parameter values in Eq. (\ref{fit}) leads us to 
something unusual. By substituting the values of $b$ and $c$ 
in the octet mass formula, we obtain 
\begin{equation}
        \left. \begin{array}{l}
            M_{\Xi^*_{\frac{1}{2}}}= 1790\; {\rm MeV}\\    
            M_{\Sigma^*({\bf 8})}= 1785\; {\rm MeV} \\
            M_{\Lambda^*} = 1600\; {\rm MeV} \\
            M_{N^*({\bf 8})} = 1520\; {\rm MeV},  \label{mass2}
            \end{array} \right.
\end{equation}
and $M_{\Sigma^*({\bf\overline{10}})}$ at 1755 MeV which is 30
MeV below $M_{\Sigma^*({\bf 8})}$. The large $\Lambda^*$-
$\Sigma^*({\bf 8})$ mass splitting and the near degeneracy of 
$\Sigma^*({\bf 8})$ and $\Xi^*_{\frac{1}{2}}$, though allowed by group 
theory, would not look natural in the constituent quark model.

\subsection{Adding a singlet-octet mixing}

A comment  is in order on the $\Lambda^*$ states. There is another
``three-star'' $\Lambda^*$ state of $\frac{1}{2}^+$ at 1810 MeV. 
If the antidecuplet is absent, $\Lambda^*(1810)$ would be assigned 
to the second excited $\frac{1}{2}^+$ octet with $N^*(1710)$, 
$\Sigma^*(1880)$, and $\Xi^*_{\frac{1}{2}}$ that is yet to be found. 
In the case that ${\bf \overline{10}}$ is present, the natural 
assignment of $\Lambda(1810)$ is an $SU(3)$ singlet by itself. 
Then a singlet-octet mixing occurs between two $\Lambda^*$ states 
with a mixing angle,
\begin{equation}
 \tan 2\theta_{\Lambda}= \frac{2\delta_{\Lambda}}{M_1 - M_8}.
\end{equation} 
The mass eigenvalues after the mixing are expressed as
\begin{eqnarray}
   \Lambda^*(1810) + \Lambda^*(1600) &=& M_1 + M_8, \nonumber \\
   (\Lambda^*(1810) -\Lambda^*(1600))\cos 2\theta_{\Lambda}
    &=& M_1 - M_8.        
\end{eqnarray}
Irrespective of the value of $\theta_{\Lambda}$, we have
\begin{equation}
              1600\;{\rm MeV} \leq M_8 \leq 1810\;{\rm MeV}, 
\end{equation}
since two levels repel each other by mixing. 
This creates a room only to raise the value of $M_8$ keeping the
$\Sigma^*({\bf 8})$ mass unchanged so that the octet mass spectrum
is more in line with the constituent quark picture. However, the
value of $\theta_N$ can be determined without knowledge of the 
$\Lambda^*$ sector since 
\begin{eqnarray}
  \tan 2\theta_N &=& \frac{2\delta}{M_{N^*(1710)}+
  M_{N^*(1440)}-2(M_{\overline{10}}-a)}, \nonumber \\
 \delta^2 &=& (M_{N^*(1710)}+M_{N^*(1440)})(M_{\overline{10}}-a)
       +(M_{\overline{10}}-a)^2 - M_{N^*(1710)}M_{N^*(1440)}, 
\end{eqnarray}
where $M_{\overline{10}}$ and $a$ are determined by the decuplet 
alone. For the same reason, $\theta_{\Sigma}$ is also not affected
by the ${\bf 1}$-${\bf 8}$ mixing.

{\subsection{Experimental uncertainties}

   A difficult question is what uncertainties should be attached to 
the values that we have obtained by use of the central values or 
the most likely values for the masses. In contrast to the masses of 
the lower baryon resonances that are determined directly with the  
peaks in cross sections, the higher resonance masses are subject 
to the systematic uncertainties involved in the methods of 
partial-wave analysis. We expect little chance of improvement in 
the experimental uncertainties in the foreseeable future.  We now 
explore how much our results would be affected by the dominant 
uncertainties.

  In the $2\times 2$ mass matrix of 
$N^*({\bf 8})$-$N^*({\bf\overline{10}})$, the diagonal entry of 
$N^*({\bf \overline{10}})$ is accurately known from the $\Theta^+$
and $\Xi^*$ masses. Then we have two unknowns left in the mass 
matrix.  The mixing angle is determined according to Eqs. 
(\ref{fit}) and (\ref{angle}) by
\begin{equation}
   \cos 2\theta_N =\frac{2\times 1647\; {\rm MeV}-M_{N^*(1710)}
             -M_{N^*(1440)}}{M_{N^*(1710)}-M_{N^*(1440)}}, 
\end{equation}
where 1647 MeV is the $N^*({\bf\overline{10}})$ mass. It is easy to 
see that the minimum mixing (the largest $\cos 2\theta_N$) occurs when 
$N^*(1710)$ is at its mimimum value, while the maximum mixing occurs 
when $N^*(1710)$ is at its maximum value. Sweeping the value of
$N^*(1440)$ over the likely range of 1430$\sim$1470 MeV\cite{PDG}, 
we find the allowed range of values for the mixing angle $|\theta_N|$,
\begin{equation}
         0.15 \leq \tan^2\theta_N^{(m)} \leq 0.53.  \label{theta}
\end{equation}
In terms of the mixing angle, Eq. (\ref{theta}) corresponds to
$21^{\circ}<|\theta_N^{(m)}|<36^{\circ}$. The minimum mixing 
$\theta_N^{(m)}=21^{\circ}$ would shift the pure antidecuplet 
mass value 1647 MeV upward by 33 MeV to the lowest edge of the 
likely mass range of $N^*(1710)$.  The large mixing angle of the 
Jaffe-Wilczek model ($\tan^2\theta_N=0.5$) is only marginally 
consistent with the upper edge of the large experimental
uncertainties. The conclusion drawn from our mass spectrum analysis 
is summarized as follows: If the baryon resonance masses are within 
the ``likely ranges'' suggested by the RPP\cite{PDG}, we would have 
to question validity of the high-precision prediction of the 
$\Theta^+$ mass by the chiral soliton model. It should be added
before concluding this section that within the chiral soliton 
model a critical review\cite{W} was presented for the precision 
and the robustness of the mass prediction. By now many chiral 
soliton advocates seem to admit an uncertainty larger than 
claimed in some of the earlier papers.

\section{Hadronic decay modes}

  The two-body hadronic decay branching fractions of $N^*(1710)$ 
are listed by RRP\cite{PDG} as
\begin{equation}
    N^*(1710)\rightarrow \left\{ \begin{array}{ll}
               N\pi,  & 10\sim 20 \% \\
               N\eta, & 6.0 \pm 1.0 \% \\
               \Lambda K, & 5\sim 25 \% \\
               \Delta\pi, & 15\sim 40 \% \\
               N\rho,  & 5\sim 25 \%.
             \end{array}  \right.    \label{decay}         
\end{equation}
We should notice here among others that the decay $N^*(1710)\to
\Delta\pi$ occurs as strongly as $N^*(1710)\rightarrow N\pi$
in spite of the smaller phase space. This is a clear evidence  
for the ${\bf 8}$-${\bf\overline{10}}$ mixing. If $N^*(1710)$ 
were a pure ${\bf \overline{10}}$, the decay 
$N^*(1710)\rightarrow \Delta\pi$ would be 
$SU(3)$-forbidden\footnote{Relevance of this selection rule was 
pointed out in \cite{Oh} and its consequence particularly in the 
decay $\Xi_{\frac{3}{2}}\to\Xi({\bf 10})^*\pi$ was 
studied in \cite{Dudek}.}  
\begin{equation}
       {\bf\overline{10}} \not{\!\to} {\bf 10}+{\bf 8}.
\end{equation}
 An educated guess is that $B(N^*({\bf\overline{10}})\to\Delta\pi)$ 
would be less than a tenth of $B(N^*({\bf\overline{10}})\to N\pi)$. 
Only if there exists a nonnegligible mixing of ${\bf 8}$, can 
$N^*(1710)$ decay substantially into $\Delta\pi$ through the 
octet component. When we compare the decay $N^*(1710)\to\Delta\pi$ 
with the decay $N^*(1440)\to\Delta\pi$ by separating the phase 
space factors, the reduced decay rate of $N^*(1710)\to\Delta\pi$ 
is considerably smaller than that of $N^*(1440)\to\Delta\pi$. 
This indicates that $N^*(1710)$ is dominantly ${\bf \overline{10}}$ 
while the ${\bf 8}$ component definitely exists. We can determine 
the ${\bf 8}$-${\bf\overline{10}}$ mixing from the partial decay 
widths into $\Delta\pi$.  Then we can compare it with the value 
required by the mass spectrum. We proceed as follows.  

     Four relevant $SU(3)$-couplings are involved in the decays of
Eq. (\ref{decay}). We denote them by
\begin{eqnarray}
  g_{8S},& g_{8A}\;\;\; {\rm for}&\;\;\; 
                          {\bf 8}\to{\bf 8}+{\bf 8},\nonumber\\
  g_{10}  \;\; &{\rm for}&\;\;\;
                      {\bf 8}\to{\bf 10}+{\bf 8},\nonumber \\
  g_{\overline{10}} \;\; &{\rm for}&\;\;\
                    {\bf\overline{10}}\to{\bf 8}+{\bf 8} 
\end{eqnarray}  
in the obvious notation. According to a simple, commonly accepted
practice, we separate the phase space factor and assume the 
$SU(3)$ symmetry holds for the dimensionless coupling constants. 
Specifically, we express the decay rates in the form of
$\Gamma = g^2(p^{2l+1}/M_i^{2l})$ where $g$ is the 
$SU(3)$-symmetric coupling, $M_i$ is the initial baryon mass, and
$l=1$ in the present case.  We shall not include the $SU(3)$ 
breaking in these coupling constants since the phase space corrections 
are by far the dominant $SU(3)$ breaking effect. Once the $SU(3)$ 
breaking were included for the coupling constants in a general form, 
we would lose simple group-theory predictions among the decay rates.
We have tabulated our 
$SU(3)$-parametrization of the decay amplitudes in Table II and
the relevant experimental numbers in Table III by defining the 
reduced width $\overline{\Gamma}$ with $\overline{\Gamma}
=(M_i^2/p^3)\Gamma$.

  While the experiment has so far given only the upper bounds on 
the $\Theta^+$ decay width, Cahn and Trilling\cite{CT} were able 
to deduce it
with a minimum of  theoretical input by combining the $\Theta^+$ 
production in $K^+Xe$ with the charge-exchange $KN$ scattering 
data at the continnum from the 1960's-70's. Since their argument 
is simple, model-independent and robust, we adopt the Cahn-Trilling 
value for $\Gamma_{\Theta^+}$;
\begin{equation}
    \Gamma(\Theta^+\to KN) = 0.9 \pm 0.3\;{\rm MeV}.  \label{CT}
\end{equation}
This constrains the value of $g_{\overline{10}}$ tightly, 
reflecting the abnormal narrowness of the $\Theta^+$ decay width:
\begin{equation}
        g_{\overline{10}}^{\;\;\;2} = 0.11\pm 0.04.  \label{tenbar}
\end{equation}
The value of Eq. (\ref{CT}) predicts the $\Xi_{\frac{3}{2}}
\to\Xi\pi$ decay width by $SU(3)$ symmetry:
$\Gamma(\Xi_{\frac{3}{2}}\to\Xi\pi)<1.9\pm 0.6$ MeV. The 
experiment\cite{CERN} has set an upper bound on the 
$\Xi_{\frac{3}{2}}$ total width at less than 18 MeV. 
 
Next we compare the $N\pi$ decay modes of $N^*(1710)$ and $N^*(1440)$.
By summing two decay rates, we obtain from Tables II and III
\begin{equation}
     \biggl(\frac{3\sqrt{5}}{10}g_{8S}+\frac{1}{2}g_{8A}\biggr)^2
          + \frac{1}{4}(g_{\overline{10}})^2 = 7.8. \label{eight}
\end{equation}
Comparing this relation with Eq. (\ref{tenbar}), we find that the
octet couplings completely dominate over the abnormally small
antidecuplet coupling. In the approximation of dropping the coupling 
$g_{\overline{10}}$ in Eq. (\ref{eight}), the mixing angle of $N^*$ 
is simply determined according to Eq. (\ref{N}) by
\begin{equation}
  \tan^2\theta_N\simeq\frac{\overline{\Gamma}(N^*(1710)\to N\pi)}{
               \overline{\Gamma}(N^*(1440)\to N\pi)}. \label{ratio}
\end{equation}
Substituting the experimental numbers in Eq. (\ref{ratio}), we 
obtain 
\begin{equation}
              \tan^2\theta_N^{(d)} \simeq 0.030,  \label{tan2}
\end{equation}
where the superscript of $\theta_N^{(d)}$ indicates the
value determined by the decays. The value of $\tan^2\theta_N^{(d)}$
turns out small despite the sizable $N^*(1710)\to\Delta\pi$ 
branching fraction. The reason is that $N^*(1710)$ is much
narrower in width than $N^*(1440)$ and that the phase 
space factor $p^3$ is much larger for $N^*(1710)$. The value of 
$\tan^2\theta_N^{(d)}$ may look qualitatively 
similar to the value or the range of values for
$\tan^2\theta_N^{(m)}$ in Eqs. (\ref{m}) or (\ref{theta}). 
However, when we compute the decay rate of $N^*(1710)\to N\pi$ 
from Eq. (\ref{ratio}) with $\tan^2\theta_N =0.15$ (cf Eq.
(\ref{theta})), the branching fraction $B(N^*(1710)\to N\pi)$ would
come out close to 100\%. That is, even the minimum value 
$\tan^2\theta_N=0.15$ allowed by the mass spectrum is much too 
large for the branching fraction $B(N^*(1710)\to N\pi)$.  
From the experimetal uncertainties quoted in RPP\cite{PDG}, 
we can put $\tan^2\theta_N^{(d)}$ in the range of
\begin{equation}
         0.008 < \tan^2\theta^{(d)} < 0.078. \label{range}
\end{equation}
The resonance $N^*(1440)$, known as the Roper resonance, 
has been best studied since the 1960's among the higher baryon 
resonances. It is difficult to reconcile the mixing angle 
$\tan^2\theta_N^{(d)}<0.08$ required by the decay with the 
value from the mass spectrum, $\tan^2\theta_N^{(m)}> 0.15$.

  Our conclusion from the decay widths is that the $\Delta\pi$
decay mode requires the presence of 
${\bf 8}$-${\bf\overline{10}}$ mixing in $N^*(1440)$ and 
$N^*(1710)$. However, the range of the values for the mixing 
angle $\theta_N^{(d)}$ required by the decays is outside 
of the corresponding range required by the mass spectrum.

  It should be noted here that very recently Praszalowicz
proposed the ${\bf\overline{10}}-{\bf 27}$ mixing\cite{Pr} in
order to improve the chiral soliton model predictions on the 
decay rates. Such a mixing would also generate the decay
$N^*(1710)\to\Delta\pi$ through the mixing. We discuss this possibility 
below in section VI.

\section{Radiative decays}

   The ${\bf \overline{10}}$ representation has a unique feature 
in the electromagentic property, as was already pointed out in
literature\cite{Close}. The electromagnetic current
is a $U$-spin singlet while the positively charged members
of ${\bf\overline{10}}$ form a $U=\frac{3}{2}$ multiplet unlike
the positively charged octet members that form a $U$-spin doublet. 
Consequently, the radiative transitions obey the $SU(3)$ 
selection rule,
\begin{equation}
  {\bf \overline{10}}(Q=+1) \not{\!\to} \gamma + 
              {\bf 8}(Q=+1).\label{s}
\end{equation}
We have some pieces of experimental information on the helicity 
amplitudes of $N^*(1440)$ and $N^*(1710)$ into $\gamma N$, 
which are listed in Table IV\cite{PDG}. Since the $SU(3)$ relations 
hold for individual amplitudes, we use the information of helicity 
amplitudes for our $SU(3)$ analysis. From the $N^{*+}\to\gamma p$ 
decay amplitudes, we obtain the mixing angle by the relation,
\begin{eqnarray}
 \tan^2\theta_N^{(r)}&=&\biggl|\frac{A_{1/2}(N^*(1710)\to\gamma p)}{
        A_{1/2}(N^*(1440)\to\gamma p) }\biggr|^2,\nonumber\\
                  &=& 0.02_{-0.02}^{+0.21}. \label{gamma}
\end{eqnarray} 
The mixing angle $\theta_N^{(r)}$ in Eq. (\ref{gamma}) again 
favors the ${\bf\overline{10}}$ dominance for $N^*(1710)$. 
It does not conflict either $\theta_N^{(m)}$ or $\theta_N^{(d)}$. 
While the uncertainty is too large to say anything more 
quantitative, it does not accommodate the Jaffe-Wilczek model. 
No useful result can be extracted from the ratio of 
$A(N^*\to\gamma n)/A(N^*\to\gamma p)$ since the $N^*\to\gamma n$ 
amplitudes involve two more $SU(3)$ constants.

  We can predict the radiative decay ratio for $\Sigma^{*+}\to
\gamma\Sigma^+$ by the selection rule of Eq. (\ref{s}) if we use 
the mixing angle $\theta_{\Sigma}$ obtained in Eq. (\ref{fit}). 
After the phase space corrections. we obtain with the value of
$\theta_{\Sigma}$ from Eq. (\ref{angle})
\begin{eqnarray}
    \frac{\overline{\Gamma}(\Sigma^{*+}(1880)\to\gamma\Sigma^+)}{
       \overline{\Gamma}(\Sigma^{*+}(1660)\to\gamma\Sigma^+} &=&
   \tan^2\theta_{\Sigma}\times \;{\rm phase}\;{\rm space}\nonumber\\
           &\simeq& 2.5.
\end{eqnarray}
Unfortunately the prospects for experimental test of this
prediction are not very promising.

\section{Prospects of octet-antidecuplet mixing}

  It is not very likely that a phase-shift analysis will
uncover another $\frac{1}{2}^+$ resonance of $Y=+1$ in the mass
region of 1650 to 1700 MeV in the near future.
Given the current data on the mass spectrum and the decay pattern,
we have ruled out zero ${\bf 8}$-${\bf\overline{10}}$ mixing
solutions. Since the presence of the Roper resonance $N^*(1440)$
has been well established, the ${\bf 8}$-${\bf\overline{10}}$
mixing between $N^*(1710)$ and the $N^*(1440)$ is the most natural
solution to the problem. Experiment is less certain for the
$\Sigma^*$ resonances. But the $\Sigma^*$ masses at 1660 MeV and
1880 MeV also show  clear ${\bf 8}$-${\bf\overline{10}}$ mixing.
The difficulty is that the
${\bf 8}$-${\bf\overline{10}}$ mixing determined from the $N^*$
masses is not consistent with the value of the mixing determined
from the $N^*$ decay modes even after the large experimental
uncertainties are taken into account.

 Is it possible that $\Xi_{\frac{3}{2}}$ and $\Theta^+$ belong to
different antidecuplets ?  Let us examine it on a purely
phenomenological basis even  though it does not fit in the
chiral soliton model nor expected in any other theoretical model.
When we inspect the mass plots of the NA49
experiment\cite{CERN}, a less prominent peak can be seen just
below 2 GeV in the $\Xi\pi$ and $\overline{\Xi}\pi$ plots. If
$\Theta^+$ and $N^*(1710)$ belong to a single unmixed antidecuplet,
their $Y=-1$ partner should be found above 2 GeV ($\simeq$ 2080 MeV).
Therefore a $\Xi_{\frac{3}{2}}$ below 2 MeV cannot be their partner.
By deducing the antidecuplet mass splitting in $Y$ from $\Theta^+$
and $N^*(1710)$, we would expect the $Y=+2$ partner of
$\Xi_{\frac{3}{2}}$ at mass below 2 MeV should be below the $KN$
threshold, that is, a stable baryon of positive stangeness.
The two-decuplet scenario is therefore ruled out provided that
the mass splitting in $Y$ of the second ${\bf\overline{10}}$ is
roughly of the same magnitude as $m_{N^*(1710)}-m_{\Theta^+}$.
If this estimate should be grossly wrong, one should look
for another $\Theta^+$.

 The argument presented above in fact holds valid even in the
case that $\Theta^+$ and $\Xi_{\frac{3}{2}}(1862)$ have
different spin-parities, say $J^P=\frac{1}{2}^+$ and
$\frac{1}{2}^-$. The same line of argument rules out most
other possibilities. How about the case that $\Theta^+$ and
$\Xi_{\frac{3}{2}}$ both carry $\frac{1}{2}^-$ ?
The spin-parity of $\frac{1}{2}^-$ is what one would
naively expect from the constituent quark model when
the goround-state pentaquarks are all in relative $s$-wave.
There is a well-established $\frac{1}{2}^-$ resonance $N^*(1650)$.
As far as the GMO mass formula is concerned, $N^*(1650)$
fits nicely to a pure ${\bf\overline{10}}$. Recall that the pure
$N({\bf \overline{10}})$ mass should be at 1647 MeV according
to Eq. (\ref{Ntenbar}).
However, the decay pattern is in variance with the antidecuplet
assignment of $N^*(1650)$: The decay branching into $N\eta$ is
too small ($0.09_{-0.06}^{+0.05})$ as compared with the $SU(3)$
prediction (0.6). The $SU(3)$-forbidden decay mode
$N^*(1650)\to\Delta\pi$ has been observed with a branching
fraction of $1\sim7$\%. Another potentially serious problem of the
$\frac{1}{2}^-$ baryons is that they decay into $\frac{1}{2}^+0^-$
in $s$-wave without the partial-wave phase space suppresion of
$l=1$. The narrow decay widths of $\Theta^+$ and $\Xi_{\frac{3}{2}}$
would become even more difficult to understand.

Within the assignment of $\frac{1}{2}^+$, the only way out of the
difficulty is to postulate that for some unknown reason the second
order SU(3) violations of $O(m_s^2)$ are unexpectedly larger than
what we have seen in the lower hadron states.  If this should be
the case, corrections to the GMO formulae would be large and the
flavor $SU(3)$ prediction on the decay branching fractions would
be subject to not only the mixing effect but also the symmetry
breaking correction to the coupling constants themselves. The
phase space correction incorporates a large SU(3) violation effect
in a specific manner. However, altering the phase space correction
does not change our qualitative conclusion. 

\section{Mixing with higher representations} 

  There is  no experimental evidence that calls for a 27-plet
or higher representation of SU(3) for baryons. Nonetheless, we should 
explore for such possibilities as well since the higher representations 
naturally appear in the chiral soliton models. 

Let us summarize the difficulty that we have encountered above. 
The $N^*(1710)\to\Delta\pi$ decay definitely requires representation 
mixing for ${\bf\overline{10}}$. Since the ${\bf\overline{10}}$ widths 
are very narrow, however, only a tiny amount of mixing is needed to 
account for the $\Delta\pi$ decay. On the other hand the mass spectrum 
deviates substantially from the equal-spacing rule of GMO for 
${\bf\overline{10}}$. The magnitude of the ${\bf 8}$-${\bf\overline{10}}$ 
mixing needed to fix this deviation is much larger than what the decay 
pattern calls for. Can we resolve this difficulty by mixing  
${\bf\overline{10}}$ with ${\bf 27}$ ? In order to make our argument 
concrete and quantitative, we assign the existing higher $\frac{1}{2}^+$ 
resonances 
of $Y=1$ and $Y=0$ to the nonexotic members of a 27-plet and see 
whether such assignment can resolve the inconsistency between
the mass spectrum and the decay pattern. With the abnormal narrowness 
of the ${\bf\overline{10}}$ decay width leading to the small mixing,
we ask whether the mass spectrum can be consistent with 
an equally small mixing in the case of ${\bf 27}$-${\bf\overline{10}}$ 
mixing.

The ${\bf 27}$-plet masses are parametrized with the same GMO formula 
as Eq. (\ref{GMO}) but with different values for 
the parameter $M_{27}$, $m_{27}$, and $m_{27}'$. 
The ${\bf 27}$-${\bf \overline{10}}$ mixing strength depends on 
$(Y,I)$ unlike the ${\bf 8}$-${\bf \overline{10}}$ mixing. We find
instead of Eq. (\ref{mixing})
\begin{equation}
   \delta_{27}\equiv
 \langle N^*({\bf 27})|N^*({\bf\overline{10}})\rangle 
 =\sqrt{\frac{3}{8}}
 \langle\Sigma^*({\bf 27},1)|\Sigma^*({\bf\overline{10}})\rangle
 =\sqrt{\frac{1}{5}}
 \langle \Xi^*({\bf 27},\frac{3}{2})|\Xi^*({\bf\overline{10}})\rangle.
                                 \label{mixing2}
\end{equation}
For a mixing as tiny as $\tan^2\theta^{(d)}\simeq 0.03$ 
as suggested by the decay widths, however, 
we may ignore the mass shifts due to mixing in a good approximation. 
So we proceed with this approximation.
Then $\Theta^+$, $N^*(1710)$, and $\Sigma^*(1880)$ fit nicely to the 
equal spacing rule with $\Delta m=$ 170 MeV. The mass of 
$\Xi^*_{\overline{10},\frac{3}{2}}$ is predicted as
\begin{equation}
    M_{\Xi^*(\overline{10})} = 2050 \;{\rm MeV}.  \label{Xitenbar}
\end{equation}
This state is certainly not the $\Xi^*(1862)$ state discovered by 
NA49\cite{CERN}.  Then, does $\Xi^*(1862)$ fit in a ${\bf 27}$ ?
Since three states, $N^*(1440)$, $\Lambda(1600)$, and $\Sigma^*(1660)$) 
are assigned to {\bf 27}, the values of all three parameters in the GMO
formula are fixed. The mass of 
$\Xi^*(1862)$ is given with the GMO formula by 
\begin{eqnarray}
  M(\Xi^*_{27,\frac{3}{2}}) 
               &=& 2M_{\Sigma^*}-M_{N^*}
                                                     \nonumber\\
                            &=& 1880 \; \mbox{\rm MeV}. 
\end{eqnarray}  
The predicted value of 1880 MeV is close enough to $\Xi^*(1862)$,
considering the uncertainty of $M_{N^*}$ and $M_{\Sigma^*}$. 
We  now have an interesting alternative of assigning $\Xi^*(1862)$
to the $I=\frac{3}{2}$ member of ${\bf 27}$ instead of 
${\bf\overline{10}}$. The mass of $\Xi^*_{27,\frac{1}{2}}$ is the same 
as in Eq. (\ref{mass2}),
\begin{equation}
      M_{\Xi^*_{27,\frac{1}{2}}} =1790\;\mbox{{\rm MeV}},
\end{equation}
provided that it does not mix much with ${\bf 8}$ or ${\bf 10}$.
The state $\Sigma_{27,2}$ of $I=2$ is exotic and does not 
mix even with ${\bf\overline{10}}$ so that its mass prediction
\begin{equation}
   M_{\Sigma_{27,2}} = 1720 \;\mbox{{\rm MeV}}
\end{equation}
is robust. The $(I=\frac{3}{2},Y=1)$ member of ${\bf 27}$ should 
be at
\begin{equation}
       M_{\Delta(27)} = 1530 \; \mbox{{\rm MeV}}.
\end{equation}  
Experimentally,  the only established $\Delta$ resonance with 
$\frac{1}{2}^+$ is at 1910 MeV, which is too far away. Although 
a $\Delta$ resonance of $\frac{1}{2}^+$ is listed at 1750 MeV
in RPP\cite{PDG}, this mass is still too far away and evidence 
is very weak for its existence
(``a one-star'' resonance). The viability of the 27-plet scenario
depends on whether  $\Xi(1862)$ decays into $\Xi^*(1530)\pi$ without
suppression and on whether a $\Delta$ resonance of 
$J^P=\frac{1}{2}^+$ will be uncovered in the neighborhood of
1530 MeV in future. In addition, a $\Sigma^{\pm}\pi^{\pm}$ 
resonance should be searched for around 1720 MeV.
The ratio of the reduced decay rates $\overline{\Gamma}
(\Xi^*(1862)\to\Xi\pi)$ and $\overline{\Gamma}(\Xi^*(1862)\to
\Sigma\overline{K})$ is unity for $\Xi^*(1862)$ of ${\bf 27}$, 
which happens to be the same as for the pure ${\bf\overline{10}}$. 
However, interference between ${\bf 27}$ and ${\bf\overline{10}}$ 
can cause departure from unity. This ratio will therefore provide 
an independent test of ${\bf 27}$-${\bf\overline{10}}$ mixing for 
$\Xi^*(1862)$. 
 
Finally a remark is in order on the mixing with ${\bf \overline{35}}$.
A state of ${\bf\overline{35}}$ cannot decay into ${\bf 8}+{\bf 8}$
nor ${\bf 10}+{\bf 8}$. Therefore this mixing would not 
explain the decay mode $N^*(1710)\to\Delta\pi$ for $N^*(1710)$ of
${\bf\overline{10}}$. If the known $\frac{1}{2}^+$ resonances are 
assigned to the nonexotic members of ${\bf\overline{35}}$, then the 
$Y=3$ (strangeness 2) member of ${\bf\overline{35}}$ would have to be 
approximately at 1000 MeV, which is totally unacceptable. In fact, 
when any one of those higher $\frac{1}{2}^+$ resonances is 
an nonexotic member of ${\bf\overline{35}}$, 
the $Y=3$ member would be most likely too light and stable against 
weak and radiative decays since its mass would be far below the decay 
threshold 1940 MeV ($KKN$). Making both the $Y=3$ and the $Y=-2$
members heavier than the $|Y|\leq 1$ members in ${\bf\overline{35}}$
is against any dynamical picture of quarks. However, one farfetched
but interesting possibility exists about ${\bf\overline{35}}$:
If we assign $\Theta^+$ and $\Xi^*(1862)$ to ${\bf\overline{35}}$,
leaving $N^*(1710)$ and $\Sigma^*(1880)$ out, the abnormal narrow
widths of $\Theta^+$ and $\Xi^*(1862)$ could be explained by the
$SU(3)$ selection rule of ${\bf\overline{35}}\not{\!\rightarrow}
{\bf 8}+{\bf 8}$ and $\not{\!\rightarrow}{\bf 10}+{\bf 8}$. The
nonresonant three-body decay into $\Xi\pi\pi$ would be the leading 
$SU(3)$-allowed decay for $\Xi^*(1862)$. Their nonexotic partners 
$N^*$ and $\Sigma^*$ are predicted at mass 1647 MeV and 1755 MeV, 
respectively, by the equal-spacing rule. Despite the interesting
feartures, we would have to resolve the problem of the ``too light
$Y=3$ state'' before we consider ${\bf\overline{35}}$ seriously.
Without more experimental input, it is not fruitful to explore 
any further along this direction.

   Our analysis uses as input the existing higher baryon resonances
that have been suggested by the phase shift analysis. When we remove 
this constraint, our analysis quickly loses its effectiveness. We 
need more experimental input before extending a scope of analysis.

\section{Summary}

  Not only do the masses of $\Theta^+$, $N^*(1710)$, and 
$\Xi^*(1862)$ fail to fit to a pure antidecuplet, but also is 
the decay mode $N^*(1710)\to\Delta\pi$ in conflict with the pure
antidecuplet assignment of $N^*(1710)$. Although the 
${\bf 8}$-${\bf\overline{10}}$ mixing can explain the failure 
of the equal spacing of the masses, one loses the basis of 
the prediction\cite{Diakonov} of the $\Theta^+$ mass from the 
$N^*(1710)$ mass. Furthermore, the mixing required for the decay 
$N^*(1710)\to\Delta\pi$ is not compatible with the mixing deduced 
from the masses.  The only viable alternative is to assign
$\Theta^+$ to ${\bf\overline{10}}$ and $\Xi^*(1862)$ to ${\bf 27}$.
In addition to the antidecuplet 
$\Xi^*_{\frac{3}{2}}$ predicted at 2050 MeV,
existence of $\Delta(1530)$ of $\frac{1}{2}^+$ and $\Sigma(1720)$ of
$I=2$, and presence of $\Xi^*(1862)\to\Xi^*(1530)\pi$ will test 
this scenario.  

\acknowledgements

This work was supported 
in part by the Director, Office of Science, Office of
High Energy and Nuclear Physics, Division of High Energy Physics,
of the U.S.  Department of Energy under contract DE-FG03-94ER40833 and
DE-AC03-76SF00098, and in part by the National Science Foundation 
under grant PHY-0098840.

\begin{table}
\caption{The spin $\frac{1}{2}^+$ resonances. All resonance states 
carry the rating of four or three stars by the Review of Particle Physics
except for $\Sigma^*(1880)$ with two stars.}
\begin{tabular}{cccc} 
Baryons & $I$ & $Y$ & Mass (MeV) \\ \hline
$\Xi^*$   & $\frac{3}{2}$& -1 & 1862 \\
$\Sigma^*$  & 1 & 0 & 1880 \\
$\Sigma^*$  &   &   & 1660 \\
$\Lambda^*$ & 0 & 0 & 1600 \\
$ N^*$   & $\frac{1}{2}$ & +1 & 1440 \\
$N^* $   &               &    & 1710 \\
$\Theta^+$ &      0      & +2 & 1540
\end{tabular}
\label{table:1}
\end{table}

\begin{table}
\caption{The isoscalar factors for $SU(3)$ parametrization of
the decay amplitudes.  The commonly defined $F/D$ ratio is equal
to $\sqrt{5}g_{8A}/3g_{8S}$.}
\begin{tabular}{ll}
Decay amplitude & $SU(3)$ isoscalar factor \\ \hline
 $\Xi^*\to\Xi\pi$&$-\frac{1}{\sqrt{2}}g_{\overline{10}}$ \\
 $\Sigma^*({\bf 8})\to N\overline{K}$
      &$\frac{\sqrt{30}}{10}g_{8S}+\frac{\sqrt{6}}{6}g_{8A}$\\
 $\Lambda^*\to\Sigma\pi$ & $-\frac{\sqrt{15}}{5}g_{8S}$ \\
 $\Lambda^*\to N\overline{K}$ & $\frac{\sqrt{10}}{10}g_{8S} +
            \frac{\sqrt{2}}{2}g_{8A}$\\
 $N^*({\bf\overline{10}})\to N\pi$& $-\frac{1}{2}g_{\overline{10}}$\\
 $N^*({\bf 8})\to\Delta\pi$&$-\frac{2\sqrt{5}}{5}g_{10}$\\
 $N^*({\bf 8})\to N\pi$&$\frac{3\sqrt{5}}{10}g_{8S}+\frac{1}{2}g_{8A}$\\ 
 $N^*({\bf 8})\to N\eta$&$-\frac{\sqrt{5}}{10}g_{8S}+\frac{1}{2}g_{8F}$\\
 $\Theta^+\to KN$ & $ -g_{\overline{10}}$\\
\end{tabular}
\label{table:2}
\end{table}

\begin{table}
\caption{The decay modes and the reduced decay width 
$\overline{\Gamma}=\Gamma M_{B^*}^2/p^3$ for $B^*\to MB$. 
The range of the $\Sigma^*(1880)$ decay width is our estimate 
since RPP does not give one.}
\begin{tabular}{cccccc}
Baryons & $\Gamma_{{\rm tot}}$ (MeV) & Modes & Branching& 
$p^3/M_{B^*}^2$ (MeV) & $\overline{\Gamma}$\\ \hline
$\Xi^*(1862)$& $<18$ &$\to\Xi\pi$ & ? & 25.9 & $< 0.7$ \\
$\Sigma^*(1880)$&$(80\sim 200)$ & &  &  &  \\
$\Sigma^*(1660)$&100 (40$\sim$200)&$\to N\overline{K}$& 0.2$\pm$0.1&
        24.1 & $8.3\pm 4.1$ \\ 
$\Lambda^*(1600)$&150 (50$\sim$250)&$\to N\overline{K}$ &
        0.23$\pm$ 0.08& 15.8  & $2.2\pm 0.8$ \\
   &      &$\to\Sigma\pi$ & 0.35$\pm$ 0.25 & 14.8 & $3.6\pm 2.5$ \\
$N^*(1710)$& 100 (50$\sim$250)&$\to N\pi$ & 0.15$\pm$0.05
           & 69.2 & $0.22\pm 0.07$\\
           & &$\to N\eta$& 0.06$\pm$ 0.01 & 23.6 & 2.5$\pm$0.4 \\
           & &$\to\Lambda K$& 0.15$\pm$ 0.10 & 6.3 & 2.4$\pm$1.6\\
           & &$\to\Delta\pi$& 0.28$\pm$ 0.013 & 20.8 & 1.4$\pm$0.7\\
$N^*(1440)$& 350 (250$\simeq$350)&$\to N\pi$& 0.65$\pm$0.05&30.2&
             7.5$\pm$0.6 \\
           & &$\to\Delta\pi$& 0.25$\pm$ 0.05& 1.4 & 63$\pm$13 \\
$\Theta^+$ & 0.9$\pm$0.3 ($<$ 7) &$\to NK$ & 1& 8.2 & 0.11$\pm$ 0.04\\
\end{tabular}
\label{table:3}
\end{table}


\begin{table}
\caption{The radiative decay amplitudes of $N^*\to N\gamma$.}
\begin{tabular}{ccc}
Baryon & Decay mode & Amplitude (GeV$^{-1/2}$) \\ \hline
$N^*(1710)$ &$\to p\gamma$ &  +0.009$\pm$0.022\\
            &$\to n\gamma$ &  $-$0.002$\pm$0.014\\
$N^*(1440)$ &$\to p\gamma$ &  $-$0.065$\pm$0.004\\
            &$\to n\gamma$ &  +0.040$\pm$0.010\\
\end{tabular}
\label{table:4}

\end{table}

\end{document}